\documentclass[amsfonts, amssymb, amsmath, showkeys, nofootinbib, preprint,floatfix, prl, superscriptaddress]{revtex4-2}

\usepackage[colorinlistoftodos, color=green!40,prependcaption]{todonotes}
\usepackage{graphicx}
\usepackage{dcolumn}
\usepackage{bm}
\usepackage{physics}
\usepackage{float}
\usepackage{xcolor}
\usepackage{amsmath}
\usepackage{amssymb}
\usepackage{units}
\usepackage[english]{babel}
\usepackage{placeins}
\usepackage{lineno}
\usepackage{mhchem}

\usepackage{lipsum}

\begin{document}

\title{Strain-induced multiferroicity in \ce{Cr1/3NbS2}}

\author{Y. Sun}
\affiliation{Department of Physics and Department of Materials Science and Engineering, University of Washington, Seattle, Washington 98195, USA}
\affiliation {Department of Physics, University of California, Berkeley, California 94720, USA}
\affiliation {Materials Science Division, Lawrence Berkeley National Laboratory, Berkeley, California 94720, USA}

\author{Y. Ahn}
\affiliation {Argonne National Lab, Lemont, IL 60439, USA}
\affiliation {Department of Physics, University of
Michigan, Ann Arbor, MI 48109, USA}

\author{D. Sapkota}
\author{H. S. Arachchige}
\author{R. Xue}
\affiliation{Department of Physics and Astronomy, University of Tennessee, Knoxville, TN, 37996, USA}

\author{S. Mozaffari}
\affiliation {Department of Materials Science and Engineering, University of Tennessee, Knoxville, TN 37996, USA}

\author{D. G. Mandrus}
\affiliation {Department of Materials Science and Engineering, University of Tennessee, Knoxville, TN 37996, USA}
\affiliation {Materials Science Technology Division, Oak Ridge National Laboratory, Oak Ridge, TN 37831, USA}

\author{L. Zhao}
\affiliation {Department of Physics, University of
Michigan, Ann Arbor, MI 48109, USA}

\author{J. Orenstein}
\email{jworenstein@lbl.gov}
\affiliation {Department of Physics, University of California, Berkeley, California 94720, USA}
\affiliation {Materials Science Division, Lawrence Berkeley National Laboratory, Berkeley, California 94720, USA}

\author{V. Sunko}
\email{vsunko@ista.ac.at}
\affiliation {Department of Physics, University of California, Berkeley, California 94720, USA}
\affiliation {Materials Science Division, Lawrence Berkeley National Laboratory, Berkeley, California 94720, USA}
\affiliation {Institute of Science and Technology Austria, Am Campus 1, 3400 Klosterneuburg, Austria }

\begin{abstract}
Multiferroic materials, in which electric polarization and magnetic order coexist and couple, offer rich opportunities for both fundamental discovery and technology. However, multiferroicity remains rare due to conflicting electronic requirements for ferroelectricity and magnetism. One route to circumvent this challenge is to exploit the noncollinear ordering of spin cycloids, whose symmetry permits the emergence of polar order. In this work, we introduce another pathway to multiferroic order in which strain generates polarization in materials that host nonpolar spin spirals. To demonstrate this phenomenon, we chose the spin spiral in the well-studied helimagnet \ce{Cr1/3NbS2}. To detect the induced polarization, we introduce the technique of magnetoelectric birefringence (MEB), an optical probe that enables spatially-resolved and unambiguous detection of polar order. By combining MEB imaging with strain engineering, we confirm the onset of a polar vector at the magnetic transition, establishing strained \ce{Cr1/3NbS2} as a type-II multiferroic.  
\end{abstract}

\maketitle

\section{Introduction}
\begin{figure}[t]
    \centering
    \includegraphics[width=1\linewidth]{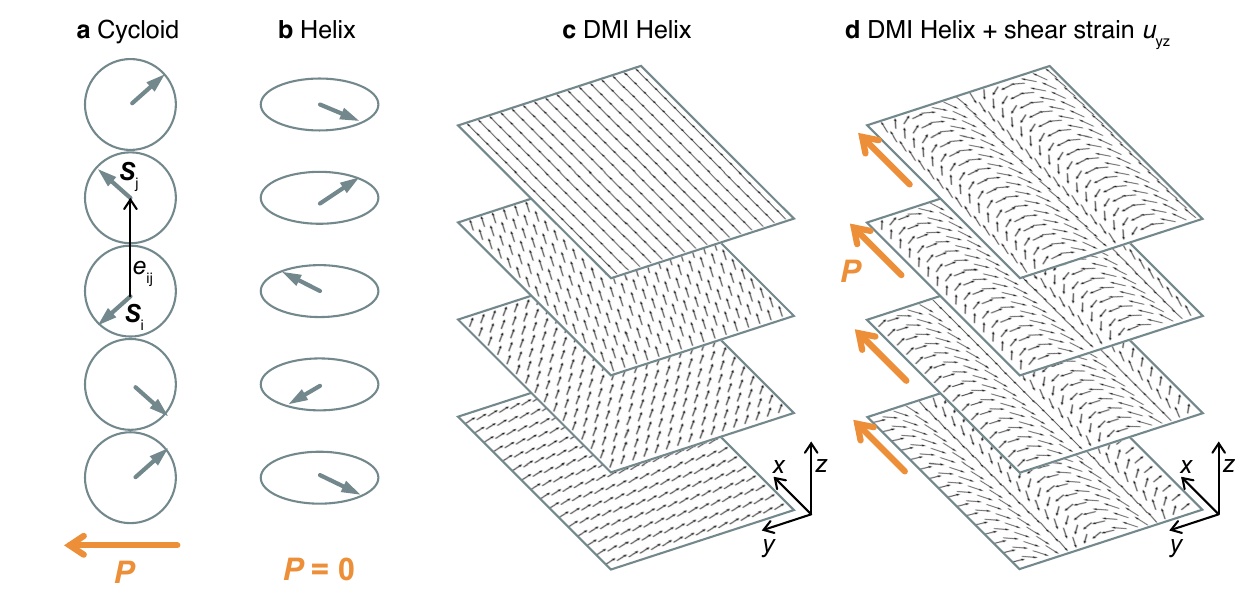}
    \caption{\textbf{Relationship between spin configuration and polarization.} Schematic of (a) spin cycloid, which gives rise to a polarization $\bm{P}$, and (b) a spin helix, which does not. (c) The helimagnetic ground state of a layered system of spins with an easy-plane anisotropy, with FM in-plane exchange and a competition between FM and DMI out-of-plane exchange, found by atomistic simulations. (d) Same as (c), but including shear strain $u_{yz}$. Helical order along the $z$ axis is maintained, but there is an additional in-plane cycloidal structure, giving rise to a polarization. The direction of polarization is set by $u_{yz}$, and is therefore the same in each plane, yielding a net polarization along $\hat{x}$. Gray and orange arrows indicate spin and polarization direction, respectively. }
    \label{fig:Fig1_spirals}
\end{figure}

Multiferroic materials, in which electric polarization and magnetic order coexist and couple, offer technologically attractive opportunities. However, intrinsic multiferroicity is rare, largely due to a ``contradiction'' in electronic requirements: ferroelectricity is favored by empty $d$-orbitals, while magnetism arises from partially filled ones \cite{spaldinAdvancesMagnetoelectricMultiferroics2019,fiebigEvolutionMultiferroics2016,mostovoyMultiferroicsDifferentRoutes2024}. One way to circumvent this contradiction is offered by noncollinear magnetic textures, which yield electric polarization $\bm{P}\propto \bm{e}_{ij}\times(\bm{S}_i\times\bm{S}_j)$ \cite{katsuraSpinCurrentMagnetoelectric2005a,mostovoyFerroelectricitySpiralMagnets2006}, with $\bm{e}_{ij}$ denoting the unit vector connecting spins  $\bm{S}_i$ and $\bm{S}_j$ (Fig.~\ref{fig:Fig1_spirals}a). Multiferroicity arising from this mechanism is classified as type-II, which means that polarization arises due to the magnetism, resulting in strong coupling between the two orders. However, not all noncollinear magnetic structures generate a net polarization \cite{spaldinAdvancesMagnetoelectricMultiferroics2019,fiebigEvolutionMultiferroics2016,mostovoyMultiferroicsDifferentRoutes2024}. Helices whose spins lie in a plane perpendicular to the propagation vector, $\bm{q}$, have $(\bm{S}_i\times\bm{S}_j)\parallel\bm{e}_{ij}$, and hence $\bm{P}=0$ (Fig.~\ref{fig:Fig1_spirals}b). 
The ability to induce polarization in such nonpolar structures has the potential to significantly expand the range of systems that stabilize emergent multiferroic phases.

Symmetry arguments suggest that shear strain applied to a nonpolar helical magnetic structure can turn it into a polar one~\cite{duTopologicalSpinStructure2021}. To illustrate this general principle, we employ atomistic simulations \cite{mullerSpiritMultifunctionalFramework2019} to find the magnetic ground state of a model Hamiltonian describing a layered system of spins with easy-plane anisotrop, ferromagnetic (FM) in-plane exchange, and a competition between FM exchange and Dzyaloshinskii–Moriya interaction (DMI) in the out-of-plane direction (see Supplementary Information~\cite{SI}). In an unstrained sample, the ground state of this Hamiltonian is a nonpolar spin-helix (Fig.~\ref{fig:Fig1_spirals}c), with $\bm{q}\parallel\hat{z}$. Adding a shear strain $u_{yz}$ induces an in-plane component of $\bm{q}$ along $\hat{y}$, and hence a cycloidal order superimposed on the zero-strain helix (Fig.~\ref{fig:Fig1_spirals}d). In this structure each layer allows for a polarization along the $\hat{x}$ direction, set by the shear strain orientation, resulting in a polar structure.   

To demonstrate strain-induced multiferroicity, we have chosen the Cr-intercalated transition metal dichalcogenides, whose ground state is a spin helix similar to that shown in Fig.~\ref{fig:Fig1_spirals}c. \ce{Cr1/3TaS2} and \ce{Cr1/3NbS2} are sister compounds belonging to the noncentrosymmetric space group P6$_3$22. Both exhibit long-wavelength helical magnetic order stabilized by DMI, with $\textbf{q}$ parallel to the crystallographic $c$-axis. In the magnetic ground state both compounds are nonpolar \cite{zhangChiralHelimagnetismOneDimensional2021,hanChiralityReversalMagnetic2023,disalvoMetalIntercalationCompounds1973,togawaChiralMagneticSoliton2012,caoOverviewAdvancesLayered2020,liuElasticallyInducedMagnetization2022}. However, it has been shown by magnetic force microscopy (MFM) that both built-in~\cite{duTopologicalSpinStructure2021} and applied~\cite{du_strain-control_2023} shear strain can significantly alter the magnetic state of \ce{Cr1/3TaS2}: an in-plane component of  $\textbf{q}$ is induced, as in our simulations (Fig.~\ref{fig:Fig1_spirals}(c,d)). The magnetic structure imaged by MFM is therefore consistent with a polar order~\cite{duTopologicalSpinStructure2021, du_strain-control_2023}. However, due to the metallic nature of these compounds, directly detecting the polar vector is challenging, and has not been achieved to date. 

Motivated by the strain sensitivity of \ce{Cr1/3TaS2}, we turn to the more extensively studied \ce{Cr1/3NbS2} ($T_c\approx$ 127 K) \cite{ghimireMagneticPhaseTransition2013,togawaInterlayerMagnetoresistanceDue2013,tsurutaPhaseDiagramChiral2016,hanTricriticalPointPhase2017} to test the prediction of polar order stabilized by shear strain in a chiral helimagnet.  As strain in as-grown crystals of \ce{Cr1/3TaS2} is inhomogeneous~\cite{duTopologicalSpinStructure2021}, this test demands spatially-resolved detection of polar order. While second-harmonic generation (SHG) is widely used as a local probe of inversion-symmetry breaking, inversion can be broken in nonpolar materials as well, making SHG an ambiguous indicator of polarization \cite{jiangDilemmaOpticalIdentification2023}.

In this work, we introduce magnetoelectric birefringence (MEB) as a sensitive and symmetry-selective probe of polar order, applicable to metallic materials. In contrast to second-harmonic generation (SHG), a nonzero MEB response unambiguously signifies the presence of a polar vector. Using MEB, we detect the emergence of a polar order parameter in \ce{Cr1/3NbS2} at the same temperature as its magnetic transition, identifying the material as a type-II multiferroic. By combining MEB imaging with substrate engineering, we further demonstrate that the polar vector is highly tunable by strain, establishing strain-induced multiferroicity in a helimagnet.

\section{Magnetoelectric birefringence: phenomenon and measurement scheme}

Magnetoelectric birefringence (MEB) was originally observed in nonpolar molecules as an optical anisotropy induced by the simultaneous application of external electric and magnetic fields~\cite{roth_observation_2002}. In solids, the same symmetry condition can be fulfilled without applied fields when spontaneous polarization $\bm{P}$ and magnetization $\bm{M}$ are present. Consequently, MEB has been observed due to spontaneous polarization~\cite{ okamura_microwave_2013, takahashi_terahertz_2013, takahashi_magnetoelectric_2012, kezsmarki_enhanced_2011, takahashi_versatile_2016, narita_observation_2016, toth_imaging_2024, kocsis_identification_2018, vit_situ_2021}. Most of these observations were made in transparent multiferroics under transmission geometries and in the terahertz range~\cite{takahashi_terahertz_2013, takahashi_magnetoelectric_2012, kezsmarki_enhanced_2011, takahashi_versatile_2016, narita_observation_2016, kocsis_identification_2018, vit_situ_2021}, where resonant effects were shown to enhance the phenomenon~\cite{takahashi_terahertz_2013, okamura_microwave_2013, takahashi_magnetoelectric_2012, kezsmarki_enhanced_2011, narita_observation_2016}. In the present work, we apply an external magnetic field to detect polar order at visible wavelengths in reflection geometry, appropriate for metallic materials, and maintaining micron-scale spatial resolution.

As was recognized in Ref.~\cite{roth_observation_2002}, in the presence of a magnetic field $\bm{H}$ and a polar vector $\bm{P}$, symmetry allows a contribution to the refractive index that is trilinear in $\bm{H}$ and $\bm{P}$, and the light wavevector $\bm{k}$,

\begin{equation}
\label{eq:MEB_def}
    n_{ME} = \left(\hat{e}\times\bm{P}\cdot \bm{k}\right)\left(\hat{e}^{*}\cdot\bm{H}\right) + \left(\hat{e}\times\bm{H}\cdot \bm{k}\right)\left(\hat{e}^{*}\cdot\bm{P}\right) + \textrm{c.c.},
\end{equation}
where $\hat{e}$ is the vector describing the polarization state of the light. 

\begin{figure}[t]
\centering
\includegraphics[width=1\columnwidth]{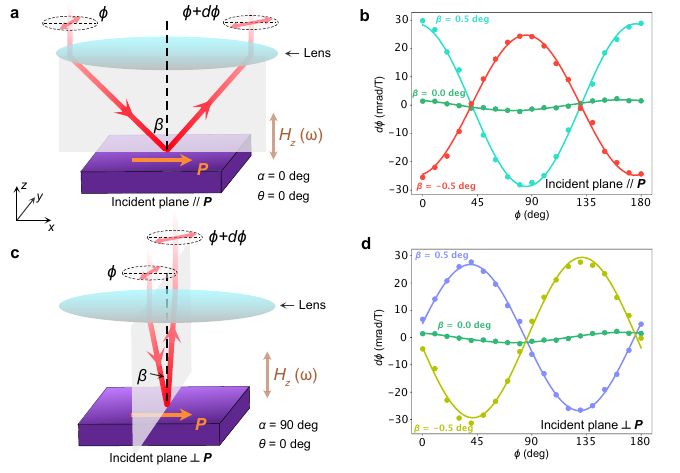}
\caption{\textbf{Experimental setup and observation of magnetoelectric birefringence.} (a) Schematic of experimental setup to measure MEB with the light-scattering plane and the polar order parameter both along the $x$-axis of the laboratory coordinate system ($\alpha=\theta = 0$ deg). (b) MEB measured in the geometry of (a) as a function of incident angle $\beta$. (c) Same as (a), but with the light-scattering plane rotated by 90 deg and perpendicular to the polar order parameter ($\alpha = 90$ deg and $\theta=0$ deg). (d) MEB measured in the geometry of (c) as a function of incident angle $\beta$. All the curves in (b) and (d) are measured under a 20 Oe oscillating magnetic field and at 10 K.}
\label{fig:MEB_observation}
\end{figure}

Here, we utilize the MEB effect to detect an in-plane polar vector, using the experimental geometry shown in Figs.~\ref{fig:MEB_observation}(a,c). We apply an out-of-plane oscillating magnetic field $H_z(\omega)$ via a copper coil and shine a laser beam with a small incident angle $\beta=\arctan{(\abs{\bm{k_{\parallel}}}/k_z)}$, shown schematically as a large angle for clarity in Fig.~\ref{fig:MEB_observation}a. 
In this geometry, and to first order in $\beta$, Eq.~\ref{eq:MEB_def} leads to birefringence (see SI~\cite{SI}), 

\begin{equation}
\label{eq:MEB_exp}
    \Delta n_{ME} = 2 \beta H_z(\omega)P_{\parallel}, \quad \varphi_\pm=\frac{\alpha-\theta \pm \pi/2}{2},
\end{equation}

\noindent where $\Delta n_{ME}$ and  $\varphi_\pm$ indicate the magnitude of birefringence and the orientation of the principal optical axes, respectively. $P_{\parallel}$ and $\theta$ are the magnitude and angle of an in-plane polar vector and $\alpha$ is the angle of the light-scattering plane (the plane determined by the incident and reflected beams). Eq.~\ref{eq:MEB_exp} indicates that $\bm{P}_{\parallel}$ can be detected if we fix the beam path ($\beta$ and $\alpha$), and measure the amplitude of birefringence that oscillates synchronously with the magnetic field $H_z(\omega)$. To do so, we measure the rotation of light polarization upon reflection ($d\phi$) as a function of incident polarization angle $\phi$, where,
\begin{equation}
\label{eq:MEB_exp_direct}
    d\phi(\omega)\propto \Delta n_{ME}\cos\left[2\phi-(\alpha-\theta)\right].
\end{equation}

The main experimental challenge in imaging the local polarization is the need to vary $\alpha$ and $\beta$ with precision, while maintaining the position of the micron-scale focal point. We realize this by employing a pair of motorized galvo-mirrors to change the incident angle while preserving beam quality (see Supplementary Information (SI) for experimental details~\cite{SI}). Images of $\bm{P}(\bm{r})$ are obtained by placing the sample on a piezoelectric stack, which translates the sample under the focused laser beam with micron scale resolution.  

\section{Polar order parameter in C\MakeLowercase{r}$_{1/3}$N\MakeLowercase{b}S$_2$}

We apply the experimental scheme described above to demonstrate the existence of polar order in \ce{Cr1/3NbS2}. First, we set the light-scattering plane along the horizontal axis of the laboratory coordinate system (this defines $\alpha = 0$), and choose a small incident angle ($\beta = 0.5^\circ$). Measuring $d\phi$ as a function of the light polarization $\phi$, we find the angular dependence predicted by Eq.~\ref{eq:MEB_exp_direct}, consistent with the presence of polar order (cyan curve in Fig.~\ref{fig:MEB_observation}b).  Changing the sign of the incident angle $\beta$, i.e. exchanging the incident and reflected beams ($\beta=-0.5^\circ$), results in the sign change of the signal (the red curve in Fig.~\ref{fig:MEB_observation}b), as predicted by Eq.~\ref{eq:MEB_exp_direct}. 
Further, we repeat the experiment with the light-scattering plane rotated by $90^\circ$, while maintaining the position of the probe focus on the sample (Fig.~\ref{fig:MEB_observation}c). As shown in Fig.~\ref{fig:MEB_observation}d, the cosine-like dependence found with $\alpha=0^\circ$ is now replaced a sine-like dependence at $\alpha=90^\circ$ in agreement with the prediction of Eq.~\ref{eq:MEB_exp_direct}. Finally, the linear dependence on $\beta$ is reproduced in this configuration as well. 

The dependence of $\Delta n_{ME}$ on both the incident angle and the plane of incidence prove that the observed phenomena are manifestations of the MEB effect, which in turn unambiguously points to the existence of polar order in \ce{Cr1/3NbS2}. Fits to Eq.~\ref{eq:MEB_exp_direct} show that the polar vector at the measurement position is oriented along the positive $\hat{x}$ direction, i.e. $\theta=0$, as shown in the schematics in Fig.~\ref{fig:MEB_observation}(a,c).

\section{Strain-induced multiferroicity}

\begin{figure}[t]
\centering
\includegraphics[width=0.5\columnwidth]{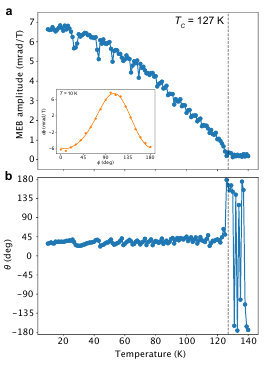}
\caption{\textbf{Temperature dependence of MEB.} Temperature dependence of (a) MEB amplitude and (b) extracted direction $\theta$ of the polar order parameter. The dashed line indicates the transition temperature of the helical spin order in \ce{Cr1/3NbS2}. Inset shows the MEB signal measured at 10 K.}
\label{fig:MEB_temp_dep}
\end{figure}

Having established the existence of polar order in \ce{Cr1/3NbS2}, we turn to studying its origin. First, we measure the temperature dependence of MEB. We set $\alpha=0^\circ$ and fix $\beta=0.5^\circ$, so the amplitude and phase of the measured sinusoidal curve directly report on the magnitude ($P_{\parallel}$) and direction ($\theta$) of the polar order parameter (\textit{c.f.} Eq.~\ref{eq:MEB_exp_direct}).  We then repeat the polarization rotation measurement varying the temperature between $\unit[5]{K}$ and $\unit[140]{K}$; an example measurement at $\unit[10]{K}$ is shown in the inset of Fig.~\ref{fig:MEB_temp_dep}a. At each temperature we fit the dependence on $\phi$ with Eq.~\ref{eq:MEB_exp_direct}, and extract the relative amplitude of the polar vector and its direction, which are shown in Fig.~\ref{fig:MEB_temp_dep}(a,b), respectively. The MEB amplitude shows a clear onset at 127 K, which coincides with the transition temperature into the helimagnetic state. This is consistent with a type II multiferroic: polar order arises simultaneously with the magnetic order. The direction of the polar order parameter does not show a strong temperature dependence, which is consistent with the hypothesis that it is determined by a built-in shear strain. However, both the  MEB amplitude and $\theta$ exhibit occasional and sudden small changes, which could reflect changes in the strain profile due to thermal contraction. 

\begin{figure}[tb]
\centering
\includegraphics[width=1\columnwidth]{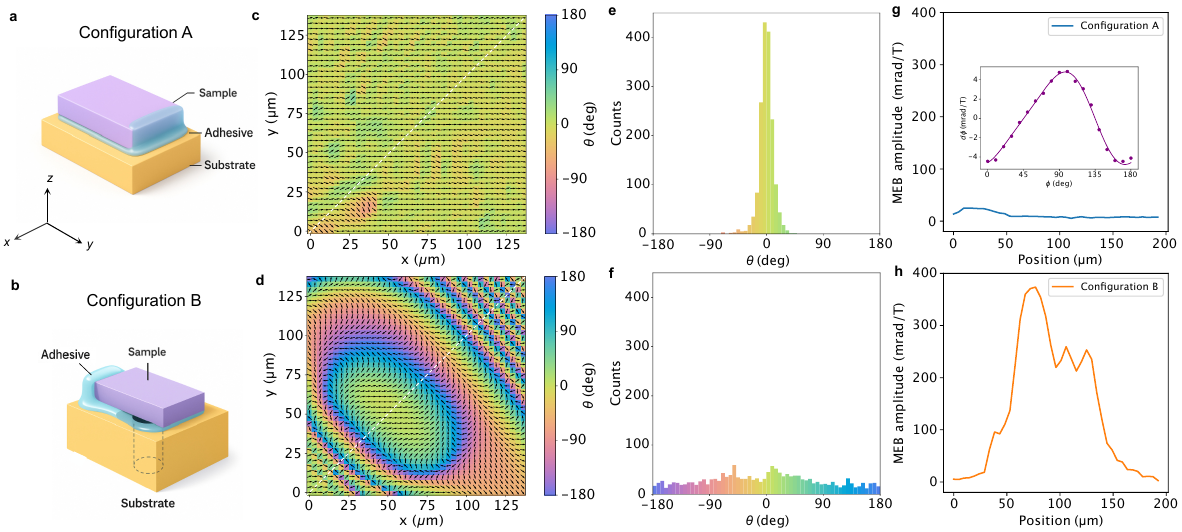}
\caption{\textbf{Strain control of the polar order parameter.} Schematic of sample mounting configuration to induce (a) uniform shear strain (configuration A) and (b) a strain distribution (configuration B), where the sample is mounted on a silicon nitride substrate with a hole. (c,d) Spatial maps of polar order parameter directions for configurations A and B respectively. The arrows indicate the direction of the polar order parameter. (e,f) Histogram of the orientations of the polar order across the maps in (c,d), respectively. (g,h) MEB amplitude along the white dashed lines in (c,d). In the inset of panel (g) we show one representative MEB measurement taken in configuration A.}
\label{fig:strain_control}
\end{figure}

To explicitly demonstrate the strain-induced nature of the observed multiferroicity, we image the spatial dependence of the MEB amplitude and $\theta$ on the same crystal mounted in two configurations, chosen to induce a uniform shear strain (configuration A) and a strain distribution (configuration B). In configuration A, shown in Fig.~\ref{fig:strain_control}a, the sample is held by an adhesive applied along the bottom surface ($xy$ plane) and along \textit{one} of the vertical sides ($zx$ plane). Thermal contraction therefore dominantly promotes the $u_{yz}$ component of shear strain. In configuration B, we rigidly mount the sample on a silicon nitride substrate with a hole (Fig.~\ref{fig:strain_control}b), inducing a highly-non-uniform strain profile. We measure MEB as a function of position in the two configurations, mapping out the same 135$\times$135 $\mu$m sample area (Figs.~\ref{fig:strain_control}(c,d); a representative MEB measurement for configuration A is shown in the inset of panel g). In configuration A, the MEB orientation is uniform, as demonstrated by the histogram of MEB orientations shown in Fig.~\ref{fig:strain_control}e. The histogram is centered at $\theta=0$, indicating a polar vector oriented along the $x$ direction, exactly as expected from the shear strain component $u_{yz}$ that we applied (\textit{c.f.} simulations in Fig.~\ref{fig:Fig1_spirals}).  

In sharp contrast, in configuration B we find a wide range of MEB angles, whose spatial profile reflects the symmetry of the hole that induced the strain (Fig.~\ref{fig:strain_control}d). The orientation distribution is very broad, almost uniformly spanning the range between $-180^\circ$ and $180^\circ$ (Fig.~\ref{fig:strain_control}f). Moreover, the MEB amplitude is on average 20 times larger in configuration B than in configuration A (Fig.~\ref{fig:strain_control}(g,h)), reflecting a larger shear strain induced by the non-uniform substrate. While reproducing the exact pattern seen in Fig.~\ref{fig:strain_control}d would require a combination of finite element analysis and microscopic magnetic simulations that is beyond the scope of this work, our findings clearly demonstrate the extreme strain-sensitivity of the polar order. With this, we confirm that type-II multiferroicity in \ce{Cr1/3NbS2} is strain-induced. 

\section{Conclusions}
We have used a new optical probe, magneto-electric birefringence, to discover strain-induced multiferroicity in \ce{Cr1/3NbS2}. This result has two significant implications for research into polar and multiferroic order. First of all, it is likely that controlled strain can be systematically used to induce multiferroicity in a broad range of non-collinear, non-polar magnetic structures. Further, the MEB measurements scheme we propose here will enable the study of polar order in regimes that are inaccessible to other tools: polar metals~\cite{bhowal_polar_2023} and two dimensional materials and engineered heterostrucures. This is particularly timely, due to numerous theoretical suggestions for emergent multiferroicity in 2D~\cite{gao_two-dimensional_2021, bennett_stacking-engineered_2024,  xu_high_2025, zhang_perspective_2024}, but a lack of appropriate optical probes that could efficiently put these theories to test. Both approaches to multiferroicity, strain engineering and reducing dimensionality are promising avenues for uncovering new multiferroic phenomena, and realizing functional devices.

\section{Acknowledgements}
Y.S., V.S. and J.O.  received support from the Gordon and Betty Moore Foundation's EPiQS Initiative through Grant GBMF4537 to J.O. at UC Berkeley. Experimental and theoretical work at LBNL and UC Berkeley was funded by the Quantum Materials (KC2202) program under the U.S. Department of Energy, Office of Science, Office of Basic Energy Sciences, Materials Sciences and Engineering Division under Contract No. DE-AC02-05CH11231. Y.S. also acknowledges support by the David J. Thouless Postdoctoral Fellowship at the Department of Physics, University of Washington. DGM acknowledges support from the Gordon and Betty Moore Foundation’s EPiQS Initiative, Grant GBMF9069. L.Z. acknowledges the support from the U.S. Department of Energy (DOE), Office of Science, Basic Energy Science (BES), under award No. DE-SC0024145

\bibstyle{apsrev4-2}

\bibliography{Main_Text_References}

\end{document}